\documentclass[aip,jmp,preprint, floatfix]{revtex4-1} 
\usepackage {graphicx} 
\usepackage {graphicx} 
\usepackage {amssymb} 
\usepackage {amsmath}
\usepackage {mathrsfs}
\usepackage{amsthm}

\newcommand{\reffig}[1]{Fig.~\ref{#1}}
\newcommand{\refeq}[1]{Eq.~(\ref{#1})}

\newcommand{\vect}[1]{\mathrm{\mathbf{#1}}} 

\newcommand{\pderiv}[2]{\partial_{#2} #1} 
\newcommand{\pdderiv}[2]{\partial_{#2 #2} #1} 

\DeclareMathOperator{\sign}{sign}

\newtheorem{theorem}{Theorem}[section]
\newtheorem{lemma}[theorem]{Lemma}

\newtheorem{definition}[theorem]{Definition}
\newtheorem{remark}[theorem]{Remark}

\newcommand{\rsp}{\mathbb{R}}
\newcommand{\csp}{\mathbb{C}}
\newcommand{\schwartz}{\mathscr{S}(\rsp^3\times \rsp)}
\newcommand{\tdist}{\mathscr{S}'(\rsp^3\times \rsp)}

\newcommand{\ft}[1]{\mathcal{F}\left[#1\right]}
\newcommand{\ift}[1]{\mathcal{F}^{-1}\left[#1\right]}
\newcommand{\fti}[2]{\mathcal{F}_{#2}\left[#1\right]}
\newcommand{\ifti}[2]{\mathcal{F}_{#2}^{-1}\left[#1\right]}

\begin{document}
\title{The fundamental solution of the unidirectional pulse propagation
  equation}

\begin{abstract}
  The fundamental solution of a variant of the three-dimensional wave
  equation known as ``unidirectional pulse propagation equation''
  (UPPE) and its paraxial approximation is obtained. It is shown that
  the fundamental solution can be presented as a projection of a
  fundamental solution of the wave equation to some functional
  subspace. We discuss the degree of equivalence of the UPPE and the
  wave equation in this respect.  In particular, we show that the
  UPPE, in contrast to the common belief, describes wave propagation
  in both longitudinal and temporal directions, and, thereby, its
  fundamental solution possesses a non-causal character.
\end{abstract}


\author{I. Babushkin} \affiliation{Institute of Mathematics, Humboldt
  University, Rudower Chaussee 25, 12489 Berlin, Germany.}
\email{ihar.babushkin@math.hu-berlin.de} \author{L. Berg\'e}
\affiliation{CEA, DAM, DIF, F-91297, Arpajon, France}

\maketitle

\section{ Introduction}
\label{sec:1}

Often in physics, the problem of light propagation in a nonlinear,
homogeneous isotropic medium requires solving the nonlinear wave
equation (WE)
\begin{equation}
  \label{eq:4}
\square E(\vect r,t) \equiv \Delta E(\vect r,t) -
 \frac{1}{c^2}\pdderiv{E(\vect r,t)}{t} = Q[E],
\end{equation}
where $\vect r = \{x,y,z\}$ is a point of  $\mathbb{R}^3$ locating the spatial coordinates, $t$ is time,
$\Delta = \partial_{zz} + \partial_{yy}+\partial_{xx}$, and $E(\vect
r,t)\in\mathbb{R}$ represents the electric field.  In the
following, we apply the scalar field assumption and suppose that the linear polarization and
nonlinearity possibly entering $Q[E]$ do not alter the polarization state. $Q[E]$ is, in general, a nonlinear operator describing the medium response.  For instance, for the case of an electromagnetic wave
propagating in a plasma, we have $Q = \mu_0 \pderiv{J}{t}$, where $J$
is the plasma current density and $\mu_0$ is the vacuum permeability. In the case
of strong optical fields, the quantity $J$ depends itself on $E$
in rather complicated way
\cite{marr67:book,berge07,couairon07,brabec08:book}, making the
equation nonlinear.

Independently on the nature of inhomogeneity $Q$, the solving for \refeq{eq:4} needs both initial and boundary conditions.
Unfortunately, this problem is, in large number of practically important cases, difficult to treat numerically. A typical situation is a
propagation of a few-cycle pulse through a waveguide
\cite{husakou01,fedotova06,babushkin07,babushkin08a,babushkin10,leblond13,demircan11,koehler11a,babushkin11,demircan12,berge13}
or in a long filament \cite{berge13,berge07}, which assumes large
extent in one spatial dimension (say, $z$), making the amount of
data required for solving the initial value problem extremely large.
To deal with such cases, Unidirectional Pulse Propagation
Equations (UPPE) have been proposed \cite{kolesik04,berge07}. One of the most well-known models aims at describing the so-called ''forward'' (propagating along positive longitudinal coordinates, $z > 0$) component of the pulse electric field in Fourier domain along the $\{x,y,t\}$ variables. It governs the Fourier-transformed electric field $\tilde E(z,\vect k_\bot,\omega)\in\mathbb{C}$ as
\begin{equation}
\label{eq:1f}
\pderiv{\tilde E(z,\vect k_\bot,\omega)}{z} -  i\beta_z \tilde E(z, \vect k_\bot,\omega) =
\frac{1}{2i\beta_z} \tilde Q(z,\vect k_\bot,\omega).
\end{equation}
Here, $\tilde E(z,k_x,k_y,\omega) = \fti{E(x,y,z,t)}{x,y,t}$, $\tilde Q =
\fti{Q}{x,y,t}$, $\fti{\cdot}{x,y,t}$ is the Fourier
transform from variables $\{x,y,t\}$ to $\{k_x,k_y,\omega\}$, namely, when using $\vect r_\bot = \{x,y\}$ and $\vect k_\bot = \{k_x,k_y\}$:
 \begin{equation}
   \fti{f}{{\vect r}_{\bot},t}{(z,{\vect k}_{\bot},\omega) = \int  f({\vect r},t) \exp{(-i {\vect k}_{\bot}  {\vect r}_{\bot} +i\omega t)}} dx dy dt,
   \label{eq:deffourier}
\end{equation} 
for a given integrable function $f$, while $\beta_z = \sqrt{\omega^2/c^2 - k_\bot^2}$. 

Equation (\ref{eq:1f}) can be easily rewritten using the original variables $\{\vect r,t\}$:
\begin{equation}
  \label{eq:1}
  \pderiv{E(\vect r,t)}{z} -  i\Phi\star E(\vect r,t) = \frac{1}{2i}\Psi \star Q(\vect r,t),  
\end{equation}
where 
\begin{equation}
\label{eq:1bis}
\Phi=\ifti{\beta_z}{k_x,k_y,\omega};\,\,\, \Psi=\ifti{1/\beta_z}{k_x,k_y,\omega},
\end{equation}
and $\star$ stands for the convolution operator with respect to the variables $\{x,y,t\}$.

In contrast to \refeq{eq:4}, \refeq{eq:1} is not PDE anymore, but it
belongs to the class of pseudo-differential equations. 
Unlike \refeq{eq:4} which is commonly integrated in time under specific conditions on boundaries and field derivatives, 
\refeq{eq:1} only requests the field value $E(x,y,t)$ at $z=0$ and 
boundary conditions in the transverse $(x,y)$ dimensions; it is then solved along the longitudinal direction $z>0$.
Equation (\ref{eq:1f}) results from a kind of factorization of the
original wave equation (\ref{eq:4}), typically \cite{kolesik04,berge07,kinsler10}
(but not necessarily \cite{amiranashvili10}) neglecting the waves
propagating backward in $z$-direction. 
{Shortly, factorization can easily be made by proceeding as follows. After applying the Fourier transform $\mathcal{F}_{\{x,y,t\}}$, \refeq{eq:4} can be written as
\begin{equation}
(\partial_z - i\beta_z)(\partial_z + i\beta_z)\tilde E = \tilde Q.
\label{uppeder}
\end{equation}
Now, we decompose the field $\tilde E$ into the
  sum $\tilde E=\tilde E_{-}e^{-i\beta_zz} + \tilde
  E_{+}e^{+i\beta_zz}$, assuming that $\tilde E_\pm$ varies in $z$ much slower
  than the exponential term, that is, $(\partial_z+i\beta_z)\tilde
  E_+e^{+i\beta_zz}\approx 2i\beta_z\tilde E_+e^{+i\beta_zz}$. The
  same decomposition into $\tilde Q_+$, $\tilde Q_-$ is made for
  $\tilde Q$. Substituting the previous quantities into \refeq{uppeder}, multiplying
  it by $e^{+i\beta_zz}$ and integrating over some short range $z$ we
  obtain $2i\beta_z(\partial_z - i\beta_z)\tilde E_+ = \tilde
  Q_+$. Now, if we assume that both the field $\tilde E$ and the
  inhomogeneity $\tilde Q$ contain \textit{only} the part
  corresponding to $\tilde E_+$ ($\tilde Q_+$), we then arrive to \refeq{eq:1f}. In this simple derivation, it is
  explicitly assumed that the amplitude $E_+$ is slow compared to
  $e^{+i\beta_zz}$. This assumption is not necessary if more
  involved projection techniques are used, which include decomposition
  of $E$ and $Q$ into some transverse modes
  \cite{kolesik04}. These derive from the linear
    modes of \refeq{eq:4}, assuming weak  nonlinearities.

Equation (\ref{eq:1f}), its 1-dimensional analogs and other modifications are
nowadays routinely utilized in optics to describe ultrashort pulses with ultrabroad spectra (see
\cite{husakou01,fedotova06,berge07,babushkin08,babushkin08a,husakou09a,husakou09,im10,amiranashvili10,
  babushkin10,babushkin10a,kolesik10a,babushkin11,koehler11a,demircan11,whalen12,andreasen12,demircan12,leblond13,berge13}
and references therein), because they include minimal assumptions
about the spectral width of $E(\vect r,t)$. A currently-met approximation is the paraxial assumption, $k_z \simeq k$, performed in the denominator of the $Q$ term in \refeq{eq:1f} \cite{berge07}, which signifies that the inhomogeneity is approached by the value defined by its forward component near the propagation axis. This approximation will be addressed at the end of the present work. Both \refeq{eq:4} and \refeq{eq:1f} can be solved analytically only in
exceptional cases. Nevertheless, if we consider the wave equation
\refeq{eq:4}, a lot can be said about the general behavior of its
solutions, considering the right hand side $Q$ as a pre-known quantity
and thus \refeq{eq:4} as a linear inhomogeneous PDE.  In particular,
in the plasma case with $Q=\mu_0 J$, it is sometimes
useful to consider an approximation in which the current $J$ does not
depend on $E$ \cite{nodland97}. Moreover, linear dispersion should in principle be treated when describing ultrashort pulses. Chromatic dispersion is embedded in the linear modes of \refeq{eq:4}, when one considers a frequency-dependent dielectric constant $\epsilon(\omega)$, and it usually intervenes through $\tilde Q = -  (\omega^2/c^2) \chi^{(1)} (\omega) \tilde E$, where $\chi^{(1)}$ denotes the first-order susceptibility tensor of the material. Accounting for noticeable variations of this quantity would, however, limit our analytic treatment. Therefore, for technical convenience, we shall consider the basic configuration in which the dielectric constant is constant.

Under these conditions, it is well-known that a solution of the linear inhomogeneous variant
of \refeq{eq:4} with a given, regular inhomogeneity $Q(\vect r,t)$ can be obtained using a
fundamental solution approach with the help of tempered distributions, i.e., the solution (in the sense of
generalized functions) is deduced from that with an inhomogeneity being a Dirac
$\delta$-function, $Q=\delta(\vect r,t)$.  The two most useful
linearly-independent fundamental solutions of the D'Alembertian operator $\square$ are 
 \begin{equation}
  \label{eq:21a}
   \mathcal{E}_{\square\pm} = \frac{-1}{4 \pi r} \delta(t \mp r/c),
 \end{equation}
 where $r= |\vect r|$.  They describe spherical waves propagating
 forward ($\mathcal{E}_{\square+}$) or backward
 ($\mathcal{E}_{\square-}$) in time. From these two solutions, only $\mathcal{E}_{\square+}$ 
 is physically meaningful, since it describes
   a response to an excitation (delta-function), that propagates in
   positive direction along time $t$ and thus respects the causality
   principle. By contrast, $\mathcal{E}_{\square-} \propto \delta(t+r/c)$ describes a response going ''backward'' in time and thus being unphysical. Therefore, any solution of \refeq{eq:4} for regular enough function $Q$ will physically make sense through the convolution product $\mathcal{E}_{\square+} \star Q$, whereas $\mathcal{E}_{\square-} \star Q$ cannot fulfill the causality principle.

   Similarly, one can search for the fundamental solution of \refeq{eq:1}, that is, its
   generalized solution using the inhomogeneity $Q=\delta(\vect r,t)$.
   To the best of our knowledge, neither such a fundamental solution,
   nor its basic properties have been investigated so far to
   appreciate the applicability of the proposed UPPE models.
   Therefore, in the present article, we construct a fundamental
   solution of \refeq{eq:1}. We
   show that this solution is a projection of the fundamental solution
   of \refeq{eq:4} to some functional subspace, formed by waves
   propagating either ``forward-'' or ``backward-'' in $z$-direction
   (see Theorem~\ref{th:1}). We explore consequences of this result
   such as the intrinsic non-causality of solutions to \refeq{eq:1}. We also
   consider a variation of the latter equation, when
     its right-hand side is stated in the paraxial
   approximation.

The paper is organized as follows. Section \ref{sec:assumpt-denot} specifies notations and definitions used in this work. Section \ref{sec:3a} defines the projecting operators and their related rules. Section \ref{sec:3} elucidates the fundamental solution of the UPPE (\ref{eq:1f}), while Section \ref{sec:4} focuses on the paraxial approximation, $k_z \simeq k$, applied to its right-hand side. Section \refeq{sec:conclusions} concludes our analysis.

\section{Assumptions and notations}
\label{sec:assumpt-denot}

Let us preliminarily introduce some notations and basic definitions. 
First, our solution will be searched in the sense of distributions, that is, we assume that all
coming functions $u(\vect r,t)$ are tempered distributions $u \in
\tdist$ belonging to the space dual to Schwartz space $\schwartz$,
with the scalar product defined as 
\begin{equation}
\label{scalprod}
\langle u,\phi\rangle = \int u\,\bar\phi \,dx dy dz dt, 
\end{equation}
where the bar symbol denotes complex conjugate. With this assumption, the Fourier transform of all
distributions considered here exists. This transform, $\mathcal{F}[u]$, satisfies $\langle\mathcal{F} u,\phi\rangle \equiv \langle
u,\mathcal{F}\phi\rangle$ \cite{vladimirov:book} and the inverse Fourier transform satisfies $\langle\mathcal{F}^{-1} u,\phi\rangle \equiv \langle
u,\mathcal{F}^{-1}\phi\rangle$.
The Fourier transform of $\phi\in\schwartz$ is defined by
 \begin{equation}
   \ft{\phi(\vect r, t)}(\vect k,\omega) = \int \phi(\vect
   r,t) \exp{(-i\vect k\vect r +i\omega t)} dx dy dz dt, \label{eq:62}
\end{equation}  
where $\vect k=\{k_x,k_y,k_z\}$, while
 \begin{equation}
   \ift{\phi(\vect k, \omega)}(\vect r, t) = 1/(2 \pi)^4 \int \phi(\vect k,\omega) \exp{(i\vect k\vect r -
     i\omega t)} dk_xdk_ydk_z d\omega. \label{eq:68}
 \end{equation}
 Note that the signs in spatial and temporal parts of the Fourier
 transform are different, following the convention used in
 electrodynamics \cite{jackson62}. As already done in \refeq{eq:deffourier}, we will also employ partial Fourier transform with respect to some subsets of variables, e.g., $\vect r_\bot = \{x,y\}$. We recall that such a partial Fourier transform of
the Dirac delta-function yields
$\fti{\delta(\vect r,t)}{\vect r_\bot, t} =
\delta(z)$. The inverse of this partial Fourier transform is defied as:
\begin{equation}
  \label{eq:3}
   \ifti{\phi(z,\vect k_\bot, \omega)}{ k_\bot,\omega}(\vect r, t) = 1/(2 \pi)^3 \int \phi(z,\vect k_\bot,\omega) \exp{(i\vect k_\bot\vect r_\bot -
     i\omega t)} dk_xdk_ydk_z d\omega.
\end{equation}
Partial Fourier transforms and their combinations are
always possible for the test functions, and hence for tempered
distributions, because each particular Fourier
transform leaves the function $\phi\in\schwartz$ in $\schwartz$.  In
this regard, we remind that Plancherel's formula applies, i.e.,
\begin{equation}
  \label{eq:16}
  \langle u,\phi\rangle = \frac{1}{(2\pi)^4}\langle\ \mathcal{F}
  u,\mathcal{F}\phi\rangle.
\end{equation}

If it exists, the convolution used in \refeq{eq:1} is defined for the distributions of $\tdist$ as 
  \begin{equation}
    \label{eq:2}
    u \star v = \mathcal{F}^{-1}_{\vect
      k_\bot,\omega}\left[\fti{u}{\vect r_\bot,t}\cdot \fti{v}{\vect r_\bot,t}\right].
  \end{equation}
  We remark that, if $u\in\tdist$, then $\Phi \star u\in\tdist$. In
  contrast, $\Psi\star u$ may not be in $\tdist$ or may even not
  exist. This happens, for instance, if $\tilde u=\mathcal{F}_{(\vect
    r_\bot,t)}[u]$ is a distribution localized at a point where
  $\beta_z=0$.

We will also employ the following definition of the Heaviside step-function:
\begin{equation}
     \label{eq:21}
     \Theta(x) = \dfrac12\left(1+\sign(x)\right).
\end{equation}
with $\Theta(0)=1/2$, since $\sign(0)=0$. We can now define the fundamental solution of \refeq{eq:1}:
 \begin{definition}
  The fundamental solution of \refeq{eq:1} is a tempered distribution
   $\mathcal{E} \in \tdist$ giving a solution (in the sense of
   distributions) of \refeq{eq:1} with $Q=\delta(\vect r,t)$.
 \end{definition}
Consequently, the solution of
 \refeq{eq:1} for an arbitrary $Q \in \tdist$ exists and is given by the
 convolution product $\mathcal{E}\star Q$, if the latter exists.
Of course, the fundamental solution is defined up to a solution of a
homogeneous problem. We will aim to find the fundamental solution which
is ''most similar'' to elementary WE solutions by \refeq{eq:21a}.

\section{Forward- and Backward- Propagating Waves}
\label{sec:3a}

Equation (\ref{eq:1}) which we consider is in some sense
''anisotropic'': The direction $z$, in contrast to the same variable entering the wave
equation (\ref{eq:4}), cannot be treated as the other
spatial coordinates. Thus, before we proceed further, we must define
the notion of ``forward-'' and ``backward-'' propagating waves with respect to the direction $z$.
First, this is done for the functions representing plane waves in the form $f(\vect r,t) = e^{i\vect k \vect
  r - i\omega t}$.  The function $f(\vect r,t)$ for arbitrary $\vect k
= \{k_x,k_y,k_z\} \in \mathbb{R}^3$, $\omega \in \mathbb{R}$, $\omega
\ne 0$, $k_z \ne 0$ is an eigenfunction of both the operator $\square$ of \refeq{eq:4} and of the operators $\Phi
\star(\cdot)$, $\Psi\star(\cdot)$ in \refeq{eq:1}. The solving for \refeq{eq:1} only needs the datum $\left. E\right|_{z=0} =
  f(z=0,x,y,t)$. From this, $f$ ``propagates'' in $z$ as $t$
increases, either backward or forward, that is, $E(z-\delta z,x,y,t) = E(z,x,y,t+\delta t)$ for
  arbitrary $\delta t\in\mathbb{R}$ and $\delta z = \omega\delta t/k_z$. From the straightforward relationship
 \begin{equation}
 \label{bwfw}
 k_z \delta z = \omega k_z \delta t/k_z^2,
 \end{equation}
 it is evident that if we change the sign of the product $\omega
k_z$, then the direction of propagation along $z$, commonly identified from the basic linear modes $\sim \mbox{e}^{\pm i k_z z}$, changes in turn. Thus, we can define the plane wave as ``forward-'' or ``backward-propagating'' in
the direction $z$ using the condition
\begin{equation}
  \label{eq:14}
    \sign k_z = \pm \sign \omega,
\end{equation}
with $+$ for forward and $-$ for backward case (see schematic representation in 
\reffig{fig:1}). 
The boundary values on the axes ($k_z=0$ or $\omega=0$) are
deliberately ascribed to both forward- and backward waves.

Being able to define the propagation direction along $z$ for a
single plane wave, we can track it down for an arbitrary combination of such
waves using the Fourier transform.

\begin{definition}
\label{def:1}
We define the projecting operators
$\hat{\mathcal{P}}_{z+}$,$\hat{\mathcal{P}}_{z-}$,
$\hat{\mathcal{P}}_{+}$,$\hat{\mathcal{P}}_{-}$,$\hat{\mathcal{P}}_{lm},
l,m=0,1$ acting from $ \tdist$ to $\tdist$ as:
  \begin{equation}
    \label{eq:45}
    \hat{\mathcal{P}}_{lm} u =
    \mathcal{F}^{-1}\left[P_{lm}(k_z,\omega)\mathcal{F}[u]\right],  
  \end{equation}
  \begin{equation}
    \label{eq:38a}
    \hat{\mathcal{P}}_{+} = \hat{\mathcal{P}}_{00} +
    \hat{\mathcal{P}}_{11}, \; \hat{\mathcal{P}}_{-} =
    \hat{\mathcal{P}}_{01} + \hat{\mathcal{P}}_{10},
  \end{equation}
  \begin{equation}
    \label{eq:38b}
    \hat{\mathcal{P}}_{z+} = \hat{\mathcal{P}}_{00} +
    \hat{\mathcal{P}}_{01}, \; \hat{\mathcal{P}}_{z-} =
    \hat{\mathcal{P}}_{10} + \hat{\mathcal{P}}_{11},
  \end{equation}
where $P_{lm}(k_z,\omega) = \Theta\left((-1)^l
    \omega\right)\Theta\left((-1)^m k_z\right) \; l,m=0,1$, $u$ is an
  arbitrary tempered distribution $ u \in \tdist $, $\Theta$ is
  defined by \refeq{eq:21}. 
\end{definition}

\begin{figure}[htbp!]
  \begin{center}
    \includegraphics[width=0.5\columnwidth]{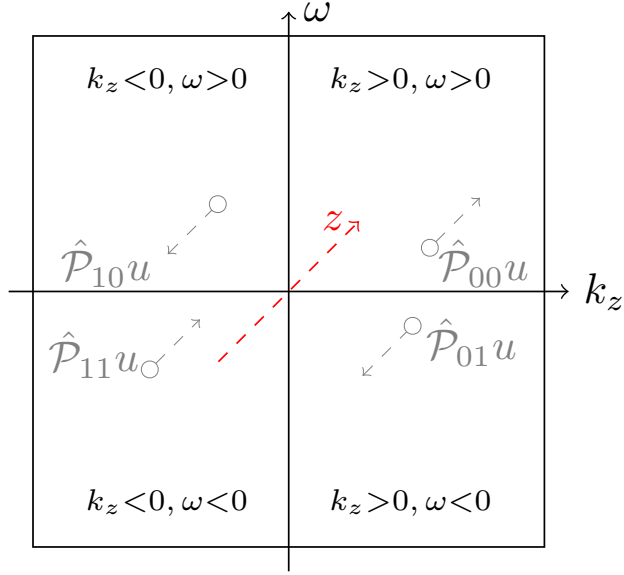}
\caption{Forward- and backward propagating waves in
  $\{k_z,\omega\}$-plane identified by the projectors
  $\hat{\mathcal{P}}_{lm}$ [\refeq{eq:45}].  Arrows parallel to
  $z$-axis show the propagation direction of each type of waves. 
  \label{fig:1}}
\end{center}
\end{figure}

The projector $\hat{\mathcal{P}}_{00}$ cuts off the part of
the spectrum of $u$ that does not belong to the quadrant
$k_z>0$, $\omega>0$ (i.e., it keeps only the upper right quadrant
in the $\{k_z,\omega\}$-plane, see \reffig{fig:1}). Similarly,
$\hat{\mathcal{P}}_{01}$ only keeps the right lower quadrant,
$\hat{\mathcal{P}}_{01}$ filters the left upper quadrant and
$\hat{\mathcal{P}}_{11}$ the left lower quadrant of the
$\{k_z,\omega\}$-plane. It is easy to see that, according to
\refeq{eq:21}, the border values $k_z=0$ or $\omega=0$ make an equal
impact to both forward- and backward-propagating
waves. This property formally follows from our
  definition of $\Theta$ in \refeq{eq:21}, which maps the point $0$ to
  $1/2$ at the frontiers of two quadrants. This is intuitively reasonable,
  because for waves propagating exactly perpendicularly to the
  $z$-axis, one cannot decide whether they propagate forward or
  backward along $z$. Obviously, all the operators defined above are continuous
and linear, and they possess the property
\begin{equation}
\hat{\mathcal{P}}_{l}^2=\hat{\mathcal{P}}_{l}, \, l=\pm, z\pm; \;\hat{\mathcal{P}}_{lm}^2=\hat{\mathcal{P}}_{lm}, \, l,m=0,1,
\label{eq:9}
\end{equation}
which is common to projecting operators.

\begin{definition}
  \label{def:2}
  A tempered distribution $ u \in \tdist$, $u\ne 0$, is called
  forward- (backward-) propagating in $z$-direction or $z$-propagating iff
  $\hat{\mathcal{P}}_{+}u = u$ ($\hat{\mathcal{P}}_{-}u = u$).
\end{definition}

Equipped with these definitions we can formulate the following result:
\begin{lemma}
  An arbitrary $u(\vect r,t) \in \tdist$ can be decomposed into the
  sum of forward- and backward- $z$-propagating functions.
\label{l:1}
\end{lemma}
From Def.~\ref{def:1} we easily see that $\hat{P}_+ + \hat{P}_- =
\hat{I}$, where $\hat I$ is a unit operator.
Thus, $u=\hat{P}_+u + \hat{P}_-u$. 

It should be noted that the decomposition defined by lemma \ref{l:1}
is not unique, since we can prescribe both forward- and backward-
directions to the components of $u$ with $\left.\ft{u}(\vect
  k,\omega)\right|_{k_z=0}$, $\left.\ft{u}(\vect
  k,\omega)\right|_{\omega=0}$. Nevertheless, for the subspace
$\{ u \in \tdist : \mathcal{F}[u]_{k_z=0}=0, \,\mathcal{F}[u]_{\omega=0}=0 \}$, the decomposition is,
indeed, unique.

\begin{definition}
  \label{def:2a}
  The action of the projectors $\hat{\mathcal{P}}_{lm}$,
  $\hat{\mathcal{P_{\pm}}}$, $\hat{\mathcal{P}}_{z\pm}$ is defined as follows: 
  $$\forall (u,\phi) \in\tdist \times \schwartz,$$
\begin{equation}
    \label{eq:7}
\langle u,\hat{\mathcal{P}}_{m}\phi\rangle =
    \langle\hat{\mathcal{P}}_{m} u,\phi\rangle,\, m=\pm,z\pm,\,
\langle u,\hat{\mathcal{P}}_{lm}\phi\rangle =
\langle\hat{\mathcal{P}}_{lm}u,\phi\rangle,\, l,m=0,1. 
  \end{equation}
\end{definition}
By virtue of the Plancherel's relation (\ref{eq:16}), we deduce that
$\hat{\mathcal{P}}_\pm$, $\hat{\mathcal{P}}_{z\pm}$,
and $\hat{\mathcal{P}}_{lm}$, applied to tempered distributions, filter the corresponding regions in Fourier domain as well.

\begin{remark}
  \label{rem:1} The fundamental solutions $\mathcal{E}_{\square\pm}$ of the wave equation
   (\ref{eq:21a}) contain components propagating both forward and backward in $z$.
\end{remark}
This can be seen by
performing the temporal Fourier transform of \refeq{eq:21a}, that expresses as
$-e^{\pm i\omega r/c}/(4 \pi r)$. This expression is indeed invariant
when reverting the sign of $z$, therefore its
Fourier transform must
contain components both with $k_z>0$ and $k_z<0$ for every $\omega$. 

\section{The Fundamental Solution for the UPPE (\ref{eq:1})}  
 \label{sec:3}

 Now, equipped with the definitions in Sec. \ref{sec:assumpt-denot},
 \ref{sec:3a}, we can formulate our main result:
 \begin{theorem}
   \label{th:1} The fundamental solution $\mathcal{E} \in \tdist$ of
   \refeq{eq:1} exists and can be represented as
   \begin{equation}
     \label{eq:6}
     \mathcal{E} = \Theta(z) \hat{\mathcal{P}}_{z+}\left\{
       \mathcal{E}_{\square+}+  
       \mathcal{E}_{\square-} \right\},
   \end{equation}
   where $\mathcal{E}_{\square\pm}$, $\mathcal{P}_{z+}$ and
   $\Theta(z)$ are given by \refeq{eq:21a}, \refeq{eq:38b} and
   \refeq{eq:21}, respectively.
 \end{theorem}

\begin{figure}[tbp!]
  \begin{center}
    \includegraphics[width=\columnwidth]{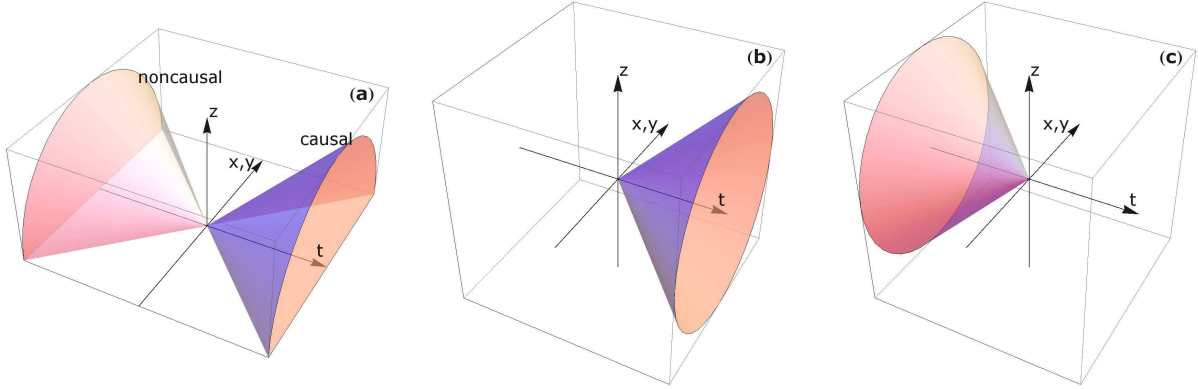}
    \caption{(a) Visual representation of the
      fundamental solution $\mathcal{E}$ defined by \refeq{eq:6}
      projected to the $\{t,x\,\mbox{or}\,y,z\}$
      plane. In (b), (c), the fundamental solutions of
        the wave equation (\ref{eq:4}), $\mathcal{E}_{\square+}$ and
        $\mathcal{E}_{\square-}$, are given for comparison [cf.
        \refeq{eq:21a}]. Only one of these, namely
        $\mathcal{E}_{\square+}$ in (b), fully respects the causality principle.
       In all other cases the waves, describing the response
        to an elementary excitation,  appear before the excitation
        itself, i.e., $\mathcal{E}$ contains components propagating both forward
        and backward in time (see the causal and noncausal zones of        
        half-cones, respectively).}
  \label{fig:2}
\end{center}
\end{figure}
 
 Before we proceed with the proof of this theorem, a few remarks are
 needed. The fundamental solution given by \refeq{eq:6} is visualized
 in \reffig{fig:2}. For $z>0,t>0$, it coincides with $\mathcal{E}_{\square+}$ and
 for $z>0,t<0$ with $\mathcal{E}_{\square-}$. Nevertheless, two
 features making it different from the fundamental solution of the wave
 equation (\ref{eq:4}) can be  immediately seen:
\begin{remark}
  In contrast to the two fundamental solutions of the wave
  equation $\mathcal{E}_{\square\pm}$, the fundamental solution
  $\mathcal{E}$ of the UPPE extends in both directions in time.
\label{rem:2}
\end{remark}
We also notice that the projector $\hat{\mathcal{P}}_{z+}$ ($\hat{\mathcal{P}}_{z-}$)
does not allow to select the set of only forward-propagating functions (resp., only backward-propagating 
functions).  Thus, taking into account the Remark~\ref{rem:1} we can conclude that:
\begin{remark}
  The fundamental solution of the UPPE (\ref{eq:6})
  contains components propagating both forward and backward in $z$.
\label{rem:3}
\end{remark}

The rest of the section is a technical proof of
Theorem~\ref{th:1}.  We proceed by solving \refeq{eq:1f}, the
Fourier-transformed analog of \refeq{eq:1} for $Q=\delta(z)$ [note that \refeq{eq:2} is written in $\{k_x,k_y,z,\omega\}$ space, so that
$\fti{\delta(\vect r,t)}{\vect r_\bot,t}=
\delta(z)$]. As it is known (see for
example \cite{vladimirov:book}) the fundamental solution of the
operator $\frac{d}{dz} + a$ for any $a\in\csp$ exists and is given by
the expression $\Theta(z) e^{-az}$. Thus, the fundamental solution
$\mathcal{E}$ of \refeq{eq:1} exists and belongs to $\tdist$, such as
\begin{equation}
  \label{eq:12}
  \mathcal{E}(\vect r,t) = \Theta(z)\ifti{\frac{\exp(i
      \beta_zz)}{2 i \beta_z}}{\vect k_\bot,\omega}.
\end{equation}

We will also use the following relation:
\begin{equation}
  \label{eq:18}
  \exp(i\beta_z z) = \lim_{\epsilon\to0}  -i \ifti{\left\{\frac{1}{k_z-\beta_z-i\epsilon} - \frac{1}{k_z-\beta_z+i\epsilon}\right\}}{k_z}. 
\end{equation}
Equation (\ref{eq:18}) must be understood in the sense of
distributions: It follows directly from Sokhotsky's
formulas $\lim_{\epsilon\to0} 1/(x\pm i\epsilon) = \mp i\pi \delta(0) + \mathscr{P}(1/x)$, where the Cauchy principal value is defined by 
\begin{equation}
  \label{eq:18bis}
  \langle \mathscr{P}(1/x),f(x) \rangle =  \lim_{\epsilon\to0} (\int_{-\infty}^{-\epsilon} + \int_{+\epsilon}^{+\infty}) \frac{f(x)}{x} dx,
\end{equation}
$\forall f\in\schwartz$. In addition, taking into account
that $k_z^2+\beta_z^2=k^2+\beta^2$ (where $\vect k = \{\vect
k_\bot,k_z\}$, $k^2=k_z^2+k_\bot^2$, $\beta = \omega/c$), the expressions
$1/(k_z-\beta_z\pm i\epsilon)$ can be rewritten in the limit $\epsilon
\rightarrow 0$ as:
\begin{equation}
  \label{eq:19}
  \frac{\pm1}{k_z-\beta_z \mp i\epsilon} = \pm
  \frac{k_z+\beta_z}{k^2-\beta^2\mp i\epsilon} = 
\pm \frac{k_z+\beta_z}{k^2-(\beta \pm i\epsilon \sign\beta)^2}.
\end{equation}
Substituting \refeq{eq:18}, \refeq{eq:19} into
\refeq{eq:12} and decomposing the resulting expression into parts
corresponding to the summands in \refeq{eq:19}, we then get
\begin{equation}
  \label{eq:39}
  \mathcal{E} \equiv \Theta(z)(\mathcal{E}_{+} + \mathcal{E}_{-});\, \mathcal{E}_{\pm} = 
  \pm \lim_{\epsilon\to0} \ift{\frac{k_z+\beta_z}{2 \beta_z} 
    \frac{1}{k^2-(\beta\mp i\epsilon\sign\beta)^2}}.
\end{equation}
We can explicitly perform the Fourier transform
$\mathcal{F}^{-1}_{\omega}$, that is, integrate over $\omega$ to obtain
\begin{equation}
  \mathcal{E}_\pm = \pm\frac{ic}{2} \mathcal{F}^{-1}_{k_x,k_y,k_z}
\left[  \Theta(k_z)\frac{e^{\mp ickt}}{k}  \right].
\label{eq:17}
\end{equation}
$\Theta(k_z)$ appears because the integration over $\omega$ gives
$\beta_z\to |k_z|$ and thus $k_z +\beta_z \to k_z + |k_z| =
2\Theta(k_z)|k_z|$. Taking into account that
$\mathcal{E}_{\square\pm}$ can be rewritten as \cite{jackson62}
\begin{equation}
  \mathcal{E}_{\square\pm} = -c\Theta(\pm t)\mathcal{F}^{-1}_{k_x,k_y,k_z}
\left[ \frac{\sin{(ckt)}}{k}  \right]
\label{eq:17a}
\end{equation}
and using \refeq{eq:39} finally leads to \refeq{eq:6}. 

\section{the UPPE with Paraxial Nonlinearity}
\label{sec:4}

The UPPE in \refeq{eq:1} is free from paraxial
approximation, i.e., no assumption is applied to the ratio
$k_z/k$ allowed in the solution. However, for practical uses,
\cite{kolesik04,babushkin10a,berge13} a simplified, but computationally more
performing variant of the UPPE may be employed, namely,
\begin{equation}
  \label{eq:55}
  \pderiv{E}{z} - i \Phi\star E = \frac{1}{2i}\Psi'\star Q,
\end{equation}
where
\begin{equation}
\Psi'=\mathcal{F}_{\{k_x,k_y,\omega\}}^{-1}[1/|\beta|],
\label{eqnew}
\end{equation}
i.e., $\beta_z$ is replaced by $|\beta|$ in the denominator of the
right-hand side of Eqs. (\ref{eq:1}), (\ref{eq:1bis}).  This equation
can be obtained using the well-known paraxial approximation
$ck_\bot/\omega\ll1$ in the inhomogeneous term of
\refeq{eq:1}. This amounts to considering that the transverse spatial components of the field are large in front of its central wavelength. Under these conditions we can omit
the diffraction wave numbers in the square root
$\beta_z = \sqrt{\omega^2/c^2 - k_\bot^2}$ affecting the inhomogeneity $Q$, thus replacing there $\beta_z$
by $|\beta|=|\omega/c|$.
\begin{definition}
  Equation (\ref{eq:55}) is called the paraxial UPPE. 
\end{definition}

\begin{theorem}
  \label{th:2}
  The fundamental solution $\mathcal{E}_{p}\in\tdist$ of
  \refeq{eq:55} exists and expresses as
\begin{equation}
  \label{eq:60}
   \mathcal{E}_{p}  = c\Theta{(z)} \int_{-\infty}^t
 \pderiv{\left\{\mathcal{E}_{-}- \mathcal{E}_{+ } \right\}}{z}\,d\tau,
\end{equation}
where $\mathcal{E}_{\pm}$ are given by \refeq{eq:39}.
\end{theorem}

Proceeding as in the previous section, we introduce
$\mathcal{E}_{p\pm}$ similarly to \refeq{eq:39}, so that

\begin{equation}
\label{eq:39pbis}
\mathcal{E}_{p}=\Theta(z)(\mathcal{E}_{p+}+\mathcal{E}_{p-}), 
\end{equation}

  \begin{equation}
  \label{eq:39p}
  \mathcal{E}_{p\pm} = \ift{ \frac{\beta_z}{|\beta|}\tilde{\mathcal{E}}_\pm},
\end{equation}
where 
 \begin{equation}
  \tilde{ \mathcal{E}}_\pm =  \frac{\pm1}{k^2-(\beta\mp
    i\epsilon\sign\beta)^2} \frac{k_z+\beta_z}{2 \beta_z}.
\label{eq:5}
\end{equation}
The formula for $\mathcal{E}_{p\pm}$ is different from \refeq{eq:39} by the factor $\beta_z/|\beta|$. Analogously to \refeq{eq:17}, the above equation can be transformed into
\begin{equation}
  \mathcal{E}_{p\pm} = \pm\frac{ic}{2} \mathcal{F}^{-1}_{k_x,k_y,k_z}
\left[ \Theta(k_z)\frac{e^{\mp ickt}}{k} \frac{k_z}{k}  \right].
\label{eq:17p}
\end{equation}
In \refeq{eq:17p} we took into account that $|k|=k$ since $k\ge0$ and $|k_z|\Theta(k_z)=k_z\Theta(k_z)$.
This equation can be re-expressed in $\{\vect r,t\}$-space as
\begin{equation}
  \label{eq:22}
  \mathcal{E}_{p\pm} =  \mp c\int_{-\infty}^t \partial_z \mathcal{E}_{\pm}\,d\tau, 
\end{equation}
which finally results in \refeq{eq:60}. 

\section{Conclusions}
\label{sec:conclusions}

In summary, although in the 3-dimensional UPPE (\ref{eq:1f}) the ``selected axis'' $z$ is pre-imposed, it enters UPPE solutions in a way, which, in many
respects, remains similar to that in the solutions to the 3-dimensional wave equation (\ref{eq:4}).
In particular, an inhomogeneity in the form of
$\delta$-function in UPPE produces wave solutions propagating both forward and backward in $z$-direction and
in time, as formulated by \refeq{eq:6},
Remarks~\ref{rem:2} and \ref{rem:3}.  This may sound in contradiction
with the commonly-used name ``unidirectional'' given to this equation.
However, if the excitation $Q$ is a
  forward- or backward-propagating function, that is $Q=\hat{
    \mathcal{P}}_{\pm}Q$, the field created by this excitation and
  given by $\mathcal{E}\star Q$ is also forward- or backward-propagating, since $\mathcal{E} \star Q = \mathcal{E} \star \hat
  {\mathcal{P}}_{\pm}Q = \hat{\mathcal{P}}_{\pm}(\mathcal{E}\star
  Q)$ by virtue of \refeq{eq:6}. Hence, reframed in its original context for which $Q$ is a nonlinear function of $E$, it is
  important that the nonlinearity $Q$ be modeled in such a way that
  it does not create backward fields.  
  
Strictly speaking, the fact that the inhomogeneity $Q$ excites
waves which propagate in both directions in time breaks the causality
principle. As illustrated in \reffig{fig:2}, the wave
  created by the excitation ($\delta$-function) can propagate both forward and backward in time, which means that
  an observer could see the result of the excitation before the latter be triggered. This is prohibited by the causality principle, and, therefore, further studies should attempt to cure this point for a better physical description of wave propagation.

  Finally, another important feature related to the number of
  spatial dimensions can also be inferred from our analysis.  By
  assuming a source created by a plasma current $J$ near some spatial
  point $\vect r_0$, the radiated field solution will be given by $E \, \sim\, Q \propto \pderiv{J}{t}$ in the framework of
  the three-dimensional ($\{x,y,z\}$) wave
  equation. Our results suggest that, up to the functional projector $\mathcal{P}_{z+}$, the
  basic proportionality $E \sim \pderiv{J}{t}$ holds for the UPPE
    (\ref{eq:1}). In contrast, the paraxial UPPE (\ref{eq:55}) would rather support the relationship $E \sim \pderiv{J}{z}$. For comparison, the 1D
  ($\{z\}$) wave equation, whose elementary solutions express as
  $\mathcal{E}_{\square\pm}\propto \Theta(t\mp|z|/c)$ (see, for
  example \cite{vladimirov:book}), would promote a radiated field solution proportional to $J$ directly,
i.e., $E \sim J$,
 since the Heaviside function assures a pre-integration in time ($\partial_t \Theta = \delta$).
\acknowledgments    

I.B. is thankful to S. Skupin, J. Herrmann and
Sh. Amiranashvili for useful discussions and to Deutsche
Forschungsgemeinschaft (DFG) for the financial support (project BA 41561-1). 


%

 \end{document}